\documentclass[aps,twocolumn,showpacs]{revtex4}
\usepackage{amsmath}%»¨Ìå×Öĸ

\usepackage{graphicx}% Include figure files
\usepackage{dcolumn}% Align table columns on decimal point
\usepackage{bm}% bold math
\usepackage{color, soul}
\usepackage[utf8]{inputenc}
\usepackage[T1]{fontenc}
\usepackage{mathptmx}
\usepackage{multirow}
\usepackage{makecell}
\usepackage{setspace}
\usepackage{footnote}
\usepackage{hyperref}
\hypersetup{hypertex=true,
            colorlinks=true,
            linkcolor=blue,
            anchorcolor=blue,
            citecolor=blue}

\begin{document}
\title{Controllable generations of several nonlinear waves in optical fibers with third-order dispersion}

\author{Peng Gao$^{1,2}$}%\email{gaopeng19@163.com}
\author{Liang Duan$^{3}$}%\email{liangduan1212@163.com}
\author{Xian-Kun Yao$^{1,2,4}$}%\email{yaoxk@nwu.edu.cn}
\author{Zhan-Ying Yang$^{1,2,4}$}\email{zyyang@nwu.edu.cn}
\author{Wen-Li Yang$^{1,2,4,5}$}%\email{wlyang@nwu.edu.cn}
\address{$^1$School of Physics, Northwest University, 710069 Xi'an, China}
\address{$^2$Shaanxi Key Laboratory for Theoretical Physics Frontiers, 710069 Xi'an, China}
\address{$^3$School of Physics and Astronomy, Shanghai Jiao Tong University, Shanghai, China}
\address{$^4$Peng Huanwu Center for Fundamental Theory, 710127 Xi'an, China}
\address{$^5$Institute of Modern Physics, Northwest University, 710069 Xi'an, China}

 %%%%%%%%%%%%%%%%%%%%%%%%%%%%%%%%%%%%%%%%%%%%%%%%%
%\date{Mar. 1, 2018}
\begin{abstract}
We propose a method to controllably generate six kinds of nonlinear waves on continuous waves, including the one- and multi-peak solitons, the Akhmediev, Kuznetsov-Ma, and Taijiri-Watanabe breathers, and stable periodic waves.
In the nonlinear fiber system with third-order dispersion, we illustrate their generation conditions by the modified linear stability analysis, and numerically generate them from initial perturbations on continuous waves.
We implement the quantitative control over their dynamical features, including the wave type, velocity, periodicity, and localization.
Our results may provide an effective scheme for generating optical solitons on continuous waves, and it can also be applied for wave generations in other various nonlinear systems.
\end{abstract}

\maketitle

\section{Introduction}
Wave generation has recently become a subject of intense research in nonlinear fiber systems.
Successful generations of some fundamental waves were implemented, such as soliton \cite{Mollenauer-1980,Stolen-1983,Mollenauer-1983,
Emplit-1987,Krokel-1988,Weiner-1988}, breathers \cite{Dudley-2009,Kibler-2012,Kibler-2015,Xu-2019}, and Peregrine rogue wave \cite{Kibler-2010,Frisquet-2013}.
It benefits from the researches on wave's exact solutions and generation mechanisms in nonlinear Schr\"{o}dinger (NLS) model \cite{Zakharov-1972,Kuznetsov-Ma,Peregrine-1983,Akhmediev-1986,Tajiri-1998,Akhmediev-2009,
Akhmediev-2009-PRE,Ankiewicz-2011,Guo-2012,He-2013}.
This famous model describes a nonlinear fiber without high-order dispersion and nonlinear effects, and it is also remarkably effective for other diverse physical systems \cite{Agrawal-book,Kevrekidis-book,Yuen-1975,Lake-1977,Bailung-2011}.
%Its mathematical simplicity and integrability offers much convenience for the research of nonlinear wave generations.

For an ultrashort pulse, some high-order effects need to be introduced into the NLS model to ensure its effectiveness \cite{Agrawal-book}.
When the coefficients of introduced effects exhibit some proportions, the model is still integrable, such as Hirota \cite{Hirota-1973} and Sasa-Satsuma \cite{Sasa-1991} models etc \cite{Ankiewicz-2016}.
Though it is difficult to realize these models in real nonlinear fibers, the researches on them provide the guidance for real wave generations.
In these models, more kinds of nonlinear waves were presented in the forms of exact solutions (especially the waves on continuous waves), like one- and multi-peak solitons and periodic waves \cite{Ankiewicz-2010,Bandelow-2012,Tao-2012,
Chen-2013,Zhao-2014,Xu-2015,Liu-2015,Liu-2016,Zhao-2016,Ankiewicz-2017}.
Their presence undoubtedly enriches the types of wave patterns, and provides new potential applications of wave generations.

The key point is how to achieve and control the generations of these waves in real fibers.
To this end, some efforts have been made to study their generation mechanism.
By the linear stability analysis, the quantitative relation between wave generations and modulation instability in the NLS model was given \cite{Dudley-2009,Zhao-2016-JOSAB}.
It provided successful controls on the generations of Akhmediev breathers and Peregrine rogue waves.
Moreover, as high-order effects was introduced, more parameters were considered to describe the features of waves, and the control over their generations was implemented numerically in an integrable model with fourth-order effects \cite{Duan-2017,Duan-2019}.
Though these existing analyses are not enough to control the wave generations in real high-order fibers, they prompt us to focus on two key features of waves, namely, periodicity and localization.
Meanwhile, the modified linear stability analysis was presented to predict the quantitative dynamics of a perturbed continuous wave \cite{Gao-2020}, including the periodicity and localization of generated waves, so it provides the possibility to control the wave generations.

In this paper, we numerically generate six kinds of fundamental waves on continuous waves, in the NLS models with third-order dispersion.
We achieve the control over their types, periodicity, localization, and velocities by the modified linear stability analysis.
The third-order dispersion enriches the types of generated waves, like the one- and multi-peak solitons on continuous waves.
The experimental feasibility of exciting the two solitons is discussed.

\section{Modified linear stability analysis for wave features}
The optical field in nonlinear fiber can be described by the NLS equation.
Considering third-order dispersion, its dimensionless model with abnormal group velocity dispersion is denoted as \cite{Agrawal-book}
%\begin{eqnarray}
%\label{eq-model}
%i\frac{\partial\psi}{\partial z}-\frac{\beta_2}{2}\frac{\partial^2 \psi}{\partial t^2}-\frac{i\beta_3}{6}\frac{\partial^3 \psi}{\partial t^3}+\sigma |\psi|^2\psi=0,
%\end{eqnarray}
\begin{eqnarray}
\label{eq-model}
i\psi_z+\frac{1}{2}\psi_{tt} -\frac{i\beta_3}{6}\psi_{ttt}+ |\psi|^2\psi=0,
\end{eqnarray}
where $\psi(z,t)$ represents the slowly-varying complex envelope of optical field.
$z$, $t$ are respectively the evolution distance and retarded time, and the subscripts $z$ and $t$ denote the partial derivative of variable with respect to them in this paper.
$\beta_3$ is corresponding to the third-order dispersion.
Though the generation conditions of some waves have been given in some integrable models with high-order effects, difficulties still remains for non-integrable systems, such as Eq. (\ref{eq-model}).

To realize their generations, we try an initial condition with the form of
\begin{equation}
\begin{split}
\label{eq-ini0p}
\psi_p(0,t)&=[1+u(0,t)]\psi_0(0,t)\\
&=[1+(a_1e^{i\omega_p t}+a_2e^{-i\omega_p t})L(t)]\;a_0e^{i\omega_0t}.
\end{split}
\end{equation}
The initial continuous wave background $\psi_0(0,t)$ has its amplitude $a_0$ and frequency $\omega_0$.
The initial perturbation $u(0,t)$ has a localized envelope $L(t)$ and the modulated periodic wave with double frequencies $\omega_p$ and $-\omega_p$, whose amplitudes are severally $a_1$ and $a_2$ (here we assume $\omega_p>0$).
The two-frequency form of periodic wave provides convenience for feature analysis of perturbing wave.
Besides, $L(t)$ is a smooth localized function whose limit values at $|t|\rightarrow \infty$ are 0.
Here, we consider $L(t)={\rm{sech}} (\eta_p t)$, where $\eta_p>0$ scales the localization of this function.
There were many studies in integrable models showing that sech-type initial perturbations can generate fundamental waves with more standard patterns than the ones with other types of envelopes \cite{Duan-2019,Conforti-2018,Liu-2018,Liu-2019}.

Now we start the analysis on dynamical features of perturbing wave.
An perturbed continuous wave is assumed as
\begin{eqnarray}
\label{eq-0p}
\psi_p(z,t)=[1+u(z,t)]\psi_0(z,t).
\end{eqnarray}
Here, $\psi_0(z,t)$ is continuous wave solution of the equation (\ref{eq-model}) with the form of $\psi_0(z,t)=a_0\exp{(i\omega_0t-ik_0z)}$, where $a_0$, $\omega_0$, and $k_0$ are severally the amplitude, frequency, and propagating constant of background wave, and $k_0=- a_0^2+\omega_0^2/2+{\beta_3}\omega_0^3/6$.
$u(z,t)$ stands for a weak perturbing wave with the condition $|u|^2\ll 1$.
Substituting Eq. (\ref{eq-0p}) into the model (\ref{eq-model}), this condition allows us to figure out the linear equation about perturbing wave $u(z,t)$:
\begin{equation}
\begin{split}
\label{eq-ueq}
iu_z&+ a_0^2u+ a_0^2u^*+(i \omega_0+\frac{i\beta_3}{2}\omega_0^2)u_t\\
&+(\frac{1}{2}+\frac{\beta_3}{2}\omega_0)u_{tt}
+(-\frac{i\beta_3}{6})u_{ttt}=0.
\end{split}
\end{equation}
We consider the perturbing wave with the form,
\begin{eqnarray}
\label{eq-uaa}
u(z,t)=A e^{p(z,t)}+B e^{p^*(z,t)}.
\end{eqnarray}
where $p(z,t)$ be a complex function about coordinates with a general form.
$A$, $B$ are the amplitudes of two conjugate components, whose initial values are equal to $a_1$ and $a_2$, respectively.
Our method is applicable only when $A$ and $B$ is small quantities in principle, due to the condition $|u|^2\ll 1$.
However, in our some numerical simulations, their moderate values compared with background amplitude also admit our control over the wave's features.
To observe the obvious wave generations, we try to set initial perturbations comparable with the background wave, which will be illustrated in Sec. IV.
We substitute Eq. (\ref{eq-uaa}) into Eq. (\ref{eq-ueq}) and obtain a set of two homogeneous equations for $a_1$ and $a_2^*$.
This set has a nontrivial solution only when the determinant of the coefficient matrix is equal to 0.
Thus, the expression of $p_z$ is
\begin{equation}
\begin{split}
\label{eq-k}
%p_{z}^{\pm m}=\beta_2 \omega_0 p_t+\frac{1}{6}\beta_3 N\pm \frac{1}{2}\sqrt{M(4\sigma a_0^2-M)},
%p_{z}^{({\rm I})}&=- \omega_0 p_t+\frac{1}{6}\beta_3 N+ \frac{1}{2}\sqrt{M(4 a_0^2-M)},\\
p_{z}=- \omega_0 p_t+\frac{1}{6}\beta_3 N\pm \frac{1}{2}\sqrt{M(4 a_0^2-M)},
\end{split}
\end{equation}
where $M=-(p_t^2+p_{tt})(1+\beta_3 \omega_0)$ and $N=p_t^3+p_{ttt}+3p_tp_{tt}-3\omega_0^2p_t$.
Comparing Eq. (\ref{eq-uaa}) with Eq. (\ref{eq-ini0p}), one can obtain that $p(0,t)=i\omega_p t+\ln L(t)=i\omega_p t+\ln [{\rm sech}(\eta_p t)]$.
It indicates that, at the initial distance $z=0$, $p_t=i\omega_p+{L_t}/{L}$, $p_{tt}={(-L_t^2+LL_{tt})}/{L^2}$, $p_{ttt}={(2L_t^3-3LL_tL_{tt}+L^2L_{ttt})}/{L^3}$,
which can be substituted into Eq. (\ref{eq-k}) for the specific expression of $p_z$.

The wave's frequency and propagation constant can be represented by the rate-of-change of phase ${\rm Im}[p]$ with $t$ and $z$ respectively,
\begin{eqnarray}
\label{eq-omk}
\omega={\rm Im}[p_t],~~K=-{\rm Im}[p_z].
\end{eqnarray}
They severally scale the wave's periodicity at $t$ and $z$ directions.
Considering that the shape of wave envelope is related to the real part of $p$, one can describe the wave's localization at $t$ and $z$ directions by the rate-of-change of ${\rm Re}[p]$, namely,
\begin{eqnarray}
\label{eq-etg}
\eta={\rm Re}[p_t],~~G=-{\rm Re}[p_z].
\end{eqnarray}
Meanwhile, the periodicity manifests as interference fringe formed by interaction between background and perturbing waves;
the localization manifests as localized envelope on background wave.
We can respectively give the velocities of fringe and localized envelope,
\begin{equation}
\begin{split}
\label{eq-v}
V=K/\omega,\quad \Lambda=G/\eta.
\end{split}
\end{equation}

Up to now, the wave's periodicity, localization, and velocities are described by the above six functions about $t$, but they are not convenient to adjust the features of wave.
Thus, we consider the relationship between $\eta_p$ and $\eta$:
\begin{eqnarray}
\label{eq-2eta}
\lim\limits_{t \to \pm \infty} \eta = \mp \eta_p.
\end{eqnarray}
It indicates that the localization of wave envelope can be described precisely by the limit of function $\eta$ when $t$ approaches positive or negative infinity.
Therefore, we introduce a subscript ($+$ or $-$) of function $f$ to denote its limit value when $t\rightarrow +\infty$ or $t\rightarrow -\infty$ : $$f_{\pm}=\lim\limits_{t\rightarrow \pm \infty}f.$$
It provides convenient limit values to depict the six features mentioned above, namely, $\omega_{\pm}$, $\eta_{\pm}$, $K_{\pm}$, $G_{\pm}$, $V_{\pm}$ and $\Lambda_{\pm}$.
For the initial condition (\ref{eq-ini0p}), we have $\omega_{+}=\omega_{-}=\omega_p$ and $\eta_{+}=-\eta_{-}=-\eta_p$.
We usually also have $K_{+}=K_{-}$ and $G_{+}=-G_{-}$ in our used model (\ref{eq-model}), which leads to $V_{+}=V_{-}$ and $\Lambda_{+}=\Lambda_{-}$ (except for the generation of one-peak soliton mentioned below).
Meanwhile, to make $V_{\pm}$ and $\Lambda_{\pm}$ describe more accurately the propagation of whole wave, the distributions of $V$ and $\Lambda$ on $t$ need to be constant approximately.
It requires that the envelope of initial perturbation is weakly localized, namely, $\eta_p<1$.

\begin{figure}[htbp]
\centering
\includegraphics[width=86mm]{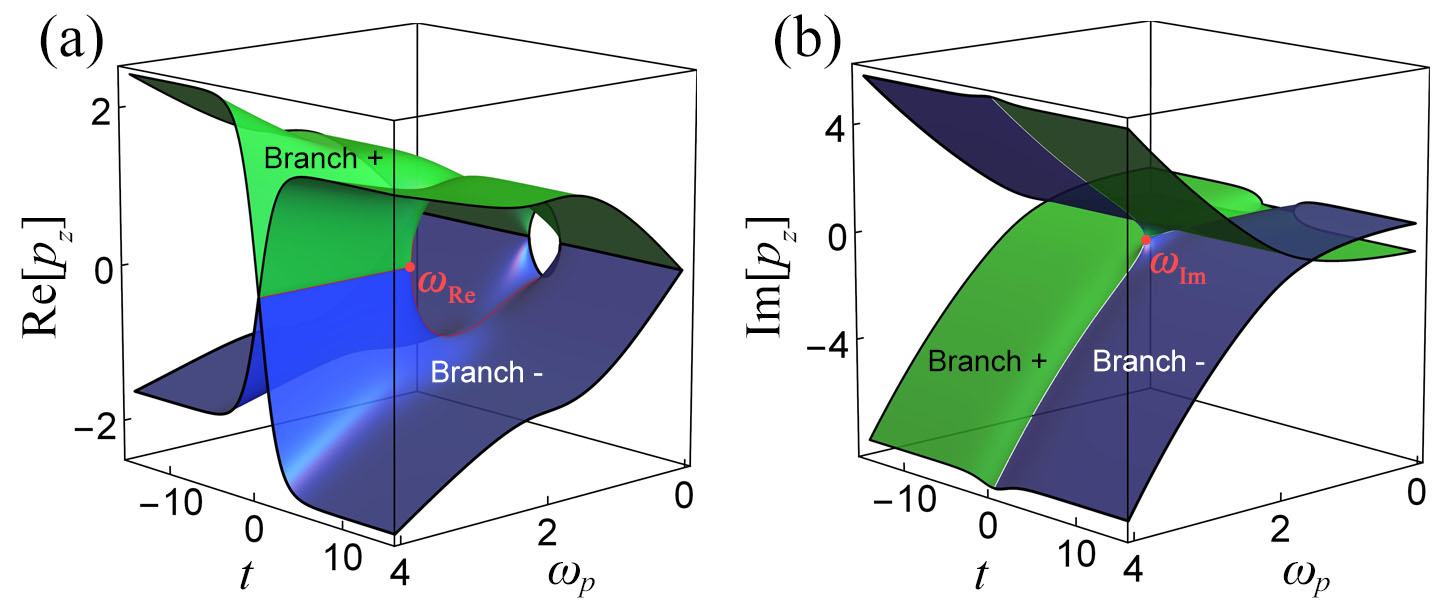}
\caption{(Color online)
Riemann surfaces of (a) ${\rm Re}[p_z]$ and (b) ${\rm Im}[p_z]$ with respect to $t$ and $\omega_p$. Branches $+$ (green surface) and $-$ (blue surface) respectively denote the function $p_z$ with $+$ and $-$ before the square-root sign. The branches of ${\rm Re}[p_z]$ (or ${\rm Im}[p_z]$) is not continuous about $t$ when $\omega_p>\omega_{\rm Re}$ (or $\omega_p>\omega_{\rm Im}$). The parameters are set as $\beta_3=0.1$, $\eta_p=0.5$, $a_0=1$, and $\omega_0=0$.}
\label{pic-rs}
\end{figure}

There exists an important problem: $p_z$ in Eq. (\ref{eq-k}) is a multi-value complex function so has two branches on its Riemann surfaces, $p_z^{\rm B+}$ and $p_z^{\rm B-}$, which indicates that the limit values of above six functions when $t\rightarrow +\infty$ and $t\rightarrow -\infty$ may locate on different branches.
For an example, when $\beta_3=0.1$, $\eta_p=0.5$, $a_0=1$, and $\omega_0=0$, the Riemann surfaces of ${\rm Re}[p_z]$ and ${\rm Im}[p_z]$ with respect to $t$ and $\omega_p$ are shown in Fig. \ref{pic-rs}.
The green or blue surfaces respectively denote the branches when the sign before the square-root sign in Eq. (\ref{eq-k}) is $+$ or $-$, called Branch $+$ or Branch $-$ here.
In Fig. \ref{pic-rs} (a), there is a red cross line between Branches $+$ and $-$, and this line ends when $\omega_p=\omega_{\rm Re}$.
In the case of $\omega_p>\omega_{\rm Re}$, the two branches are not continuous at $t=0$, so one needs to change the branch for its limit values at $t\rightarrow +\infty$ and $t\rightarrow -\infty$.
In the case of $\omega_p<\omega_{\rm Re}$, the cross line disappears so the branch is not changed.
Similarly, Fig. \ref{pic-rs} (b) shows the point where the cross line between the two branches ends, assumed as $\omega_{\rm Im}$.
The continuity about $t$ keeps in the range of $\omega_p<\omega_{\rm Im}$ and disappears when $\omega_p>\omega_{\rm Im}$.
To distinguish the two classes of results from two different continuous surfaces, we call the waves on Branch $+$ when $t\rightarrow +\infty$ Mode I, and the ones on Branch $-$ when $t\rightarrow +\infty$ Mode II.
They are respectively denoted by the superscripts (I) and (II).
The relation between the modes and branches is shown in Tab. \ref{tab-0}.
Note that this relation is applicable under the parameters in our example, which demonstrates an example way to deal with different branches.
The cases with other parameters may admit various distributions of branches so require recalculations for the relations.

\renewcommand\arraystretch{1.5}
\begin{table}[htbp]
\caption{An example relation between the modes and branches}
\label{tab-0}
\begin{tabular}{p{1.5cm}<{\centering}|p{3.2cm}<{\centering}|p{3.2cm}<{\centering}}
\hline
\hline
{\bf Cases} & {\bf Mode I} & {\bf Mode II}\\
\hline
$\omega_p<\omega_{\rm Re}$ & ${\rm Re}[p_z]^{\rm (I)}_\pm={\rm Re}[p_z]^{\rm B+}_\pm$ & ${\rm Re}[p_z]^{\rm (II)}_\pm={\rm Re}[p_z]^{\rm B-}_\pm$\\
\hline
$\omega_p>\omega_{\rm Re}$ & ${\rm Re}[p_z]^{\rm (I)}_\pm={\rm Re}[p_z]^{\rm B\pm}_\pm$ & ${\rm Re}[p_z]^{\rm (II)}_\pm={\rm Re}[p_z]^{\rm B\mp}_\pm$\\
\hline
$\omega_p<\omega_{\rm Im}$ & ${\rm Im}[p_z]^{\rm (I)}_\pm={\rm Im}[p_z]^{\rm B+}_\pm$ & ${\rm Im}[p_z]^{\rm (I)}_\pm={\rm Im}[p_z]^{\rm B-}_\pm$\\
\hline
$\omega_p>\omega_{\rm Im}$ & ${\rm Im}[p_z]^{\rm (I)}_\pm={\rm Im}[p_z]^{\rm B\pm}_\pm$ & ${\rm Im}[p_z]^{\rm (I)}_\pm={\rm Im}[p_z]^{\rm B\mp}_\pm$\\
\hline
\hline
\end{tabular}
\end{table}

As different modes can admit different dynamical features of perturbing waves, it is necessary to implement the choice of mode.
The mode choice can be realized by setting different initial amplitudes of perturbation $A$ and $B$.
With a small $B$, one can obtain the required $A$ from the homogeneous equation set,
\begin{equation}
\begin{split}
\label{eq-a1}
%a_{1}^{\pm m}={\sigma a_0^2a_2^*}/(i\beta_2\omega_0p_t+\frac{i}{6}\beta_3N+\frac{1}{2}M-ip_z^{\pm m}-\sigma a_0^2).
A^{({\rm I,II})}={ a_0^2B^*}/[-i\omega_0p_t+\frac{i}{6}\beta_3N+\frac{1}{2}M
-ip_z^{({\rm I,II})}- a_0^2].
\end{split}
\end{equation}
We know that $A^{({\rm I,II})}$ is a function with respect to $t$ and could have different limit values when $t\rightarrow +\infty$ and $t\rightarrow -\infty$.
From our experience, we need to average the values of $A_{+}^{({\rm I,II})}$ and $A_{-}^{({\rm I,II})}$, namely,
\begin{equation}
\begin{split}
\label{eq-a1ave}
%a_{1}^{(+)}=\frac{1}{2}(a_{1,t+}^{(+)}+a_{1,t-}^{(+)}),\quad
%a_{1}^{(-)}=\frac{1}{2}(a_{1,t+}^{(-)}+a_{1,t-}^{(-)}).
A_{\rm ini}^{({\rm I,II})}=\frac{1}{2}\;[A_{+}^{({\rm I,II})}+A_{-}^{({\rm I,II})}].
\end{split}
\end{equation}
It provides the required initial values of $A$ (namely $a_1$) when we generate nonlinear waves corresponding to mode ${\rm I}$ or ${\rm II}$.

By adjusting $\omega_p$, $\eta_p$, $a_0$, $\omega_0$, and $\beta_3$, the type, periodicity, localization and velocities of generated wave can be controlled.
To control the types of waves, we will classify the fundamental nonlinear waves by the above six features in the next section, and for convenience the superscript (I,II) will be omitted for a while.

\section{Nonlinear waves and their generation conditions}

\begin{figure}[htbp]
\centering
\includegraphics[width=76mm]{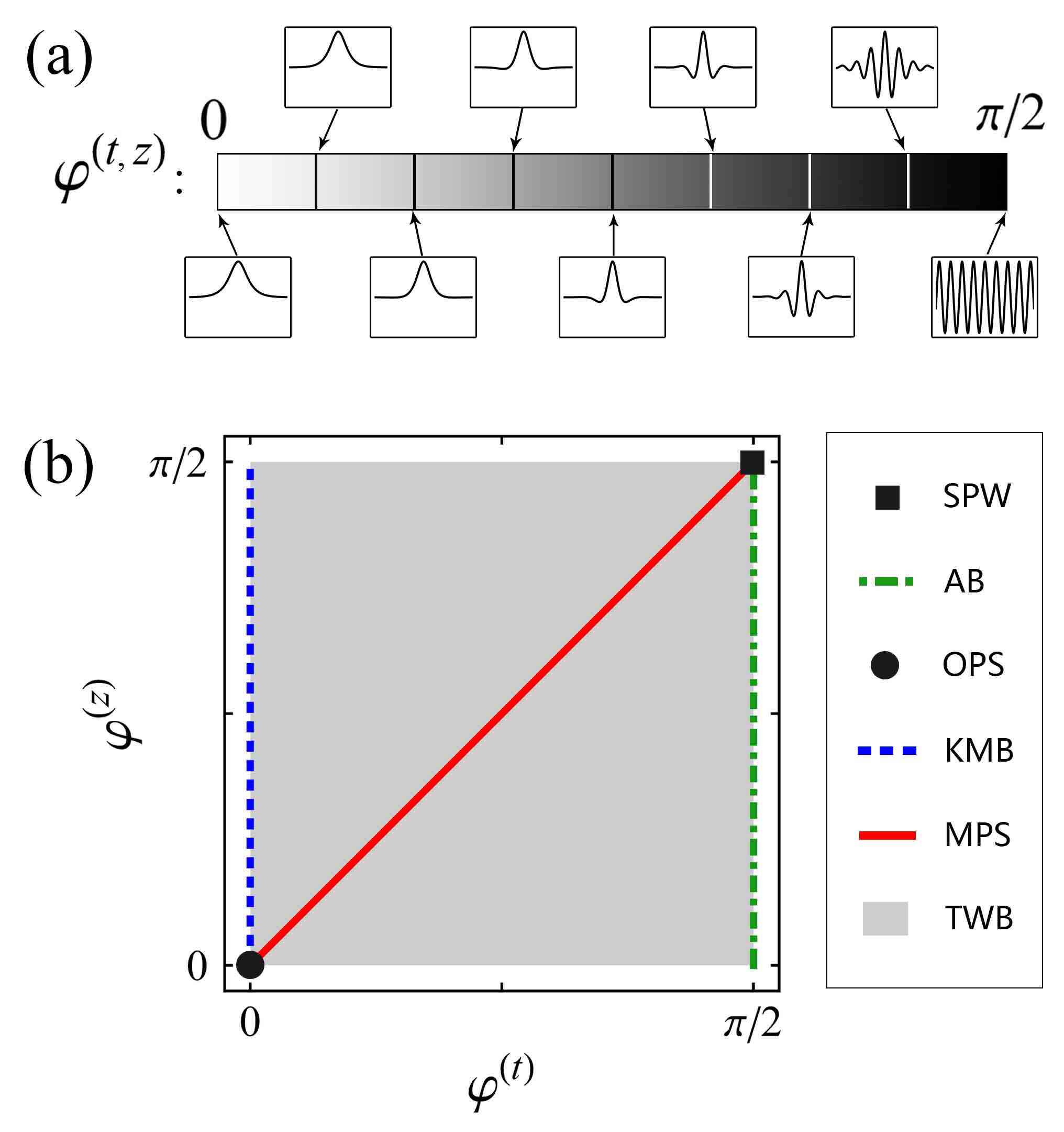}
\caption{(Color online)
(a) Typical amplitude profiles at $t$ or $z$ directions when $\varphi^{(t)}$ or $\varphi^{(z)}$ is set as different values.
The localization dominates with $\varphi^{(t,z)}$ close to $0$ while the periodicity dominates with $\varphi^{(t,z)}$ close to $\pi/2$.
(b) Feature conditions of the six kinds of waves on $\varphi^{(t)}$-$\varphi^{(z)}$ plane.
The black square, green dotted dashed line, black circle, blue dashed line, red solid line, and gray regions denote the conditions of SPW, AB, OPS, KMB, MPS, and TWB, respectively.}
\label{pic-pelo}
\end{figure}

Periodicity and localization of perturbing waves in $t$ or $z$ directions can be used to classify them, and we take the case of $t\rightarrow +\infty$ as an example here.
For a wave with the periodicity ($\omega_{+}$, $K_{+}$) and the localization ($\eta_{+}$, $G_{+}$), we define two quantities,
\begin{equation}
\begin{split}
\label{eq-phi}
\varphi^{(t)}=\tan^{-1}\Big[\frac{|\omega_{+}|}{|\eta_{+}|}\Big],\quad
\varphi^{(z)}=\tan^{-1}\Big[\frac{|K_{+}|}{|G_{+}|}\Big].
\end{split}
\end{equation}
Both of the two quantities have the range of $0\leq \varphi^{(t,z)}\leq \pi/2$ and they describe respectively the mixture of periodicity and localization in $t$ and $z$ directions.
On the $t$ or $z$ direction, the typical wave profiles corresponding to different values of $\varphi^{(t,z)}$ are shown in Fig. \ref{pic-pelo} (a).
There are three types of wave profiles:
when $\varphi^{(t,z)}=0$, the wave is purely localized;
when $\varphi^{(t,z)}=\pi/2$, the wave is purely periodic;
when $0<\varphi^{(t,z)}<\pi/2$, the wave is periodic-localized.
With the increase of $\varphi^{(t,z)}$, the wave's periodicity gets strong and the localization gets weak gradually.
The three types can help us to classify the fundamental waves.

Their classification has been discussed in some integrable models with high-order effects \cite{Liu-2015,Zhao-2016,Duan-2017}.
Here, we slightly change it and classify them into six kinds of waves, whose feature conditions are shown in Fig. \ref{pic-pelo} (b).
The six waves are discussed below.

\begin{itemize}

\item Stable periodic wave (SPW) is purely periodic in both $t$ and $z$ directions, namely, $\varphi^{(t)}=\varphi^{(z)}=\pi/2$.
\item Akhmediev breather (AB) is purely periodic in $t$ direction and is not purely periodic in $z$ direction, namely, $\varphi^{(t)}=\pi/2$ and $\varphi^{(z)}\neq \pi/2$.
\item One-peak soliton (OPS) is purely localized in both $t$ and $z$ directions, namely, $\varphi^{(t)}=\varphi^{(z)}=0$.
\item Kuznetsov-Ma breather (KMB) is purely localized in $t$ direction and is not purely localized in $z$ direction, namely, $\varphi^{(t)}=0$ and $\varphi^{(z)}\neq 0$.
\item Multi-peak soliton (MPS) is periodic-localized in both $t$ and $z$ directions. It requires that the velocities of fringe and localized envelope are the same, i.e., $V_{+}= \Lambda_{+}$, to make the wave shape stable in the evolution process. This condition of velocity matching is equivalent to $\varphi^{(t)}=\varphi^{(z)}$.
\item Taijiri-Watanabe breather (TWB) is periodic-localized in $t$ direction. To ensure the existence of breathing behavior, the velocity matching needs to be avoided, namely, $V_{+}\neq \Lambda_{+}$. It is equivalent to $\varphi^{(t)}\neq\varphi^{(z)}$.

\end{itemize}

Note that Peregrine rogue wave is not within the scope of our discussion though it has double localization in $t$ and $z$ directions.
It is because rogue wave is a special wave as the limit of breathers at $\omega_{+}, \eta_{+}, K_{+}, G_{+} \rightarrow 0$, and this limit operation leads to a rational form of waves, which cannot be described by our assumed perturbing wave with exponential form.
In a similar way, doubly periodic wave generation cannot be analyzed by our method due to its complex elliptic form.
Besides, the above feature conditions of waves are based on the analysis of periodicity and localization, which is helpful for the wave classification.
It neglects the waves with some special features, namely, the SPW with $V_{+}=0$, the OPS with $\Lambda_{+}=0$, the MPS and TWB with $V_{+}=0$ or $\Lambda_{+}=0$.
These neglected waves will be considered in the following discussion about the generation conditions of waves.

\renewcommand\arraystretch{1.5}
\begin{table}[htbp]
\caption{Generation conditions of fundamental waves}
\label{tab-1}
\begin{tabular}{p{4.7cm}<{\centering}|p{3.5cm}<{\centering}}
\hline
\hline
{\bf Fundamental waves} & {\bf Generation conditions} \\
\hline
Stable periodic wave (SPW) & $\omega_{+}\neq 0$, $\eta_{+}=0$, $G_{+}=0$\\
\hline
Akhmediev breather (AB) & $\omega_{+}\neq 0$, $\eta_{+}= 0$, $G_{+}\neq 0$\\
\hline
One-peak soliton (OPS) & $\omega_{+}=0$, $\eta_{+}\neq 0$, $K_{+}=0$ \\
\hline
Kuznetsov-Ma breather (KMB) & $\omega_{+}=0$, $\eta_{+}\neq 0$, $K_{+}\neq 0$ \\
\hline
Multi-peak soliton (MPS) & $\omega_{+}\neq 0$, $\eta_{+}\neq 0$, $V_{+}=\Lambda_{+}$\\
\hline
Taijiri-Watanabe breather (TWB) & $\omega_{+}\neq 0$, $\eta_{+}\neq 0$, $V_{+}\neq\Lambda_{+}$\\
\hline
\hline
\end{tabular}
\end{table}

Let us transform the feature conditions of the six waves into their generation conditions.
The generation conditions are decided by $\omega_{+}$, $K_{+}$, $\eta_{+}$, $G_{+}$, $V_{+}$, and $\Lambda_{+}$ to control the type and features of generated wave and are shown in the Table \ref{tab-1}.
SPW and AB can be generated from a purely periodic initial perturbation (namely $\omega_p\neq 0$ and $\eta_p=0$), generations of which can be distinguished by whether localization exists in $z$ direction or not.
In similar ways, OPS and KMB can be generated from a purely localized initial perturbation (namely $\omega_p=0$ and $\eta_p\neq 0$), generations of which can be distinguished by whether periodicity exists in $z$ direction or not;
MPS and TWB can be generated from a localized-periodic initial perturbation (namely $\omega_p\neq 0$ and $\eta_p\neq 0$), generations of which can be distinguished by whether fringes an localized envelope have identical velocity or not.
According to these generation conditions, the controllable generations of waves in the fiber system will be discussed in the next section.

\section{Controllable generation of nonlinear waves}

\begin{figure*}[htbp]
\centering
\includegraphics[width=120mm]{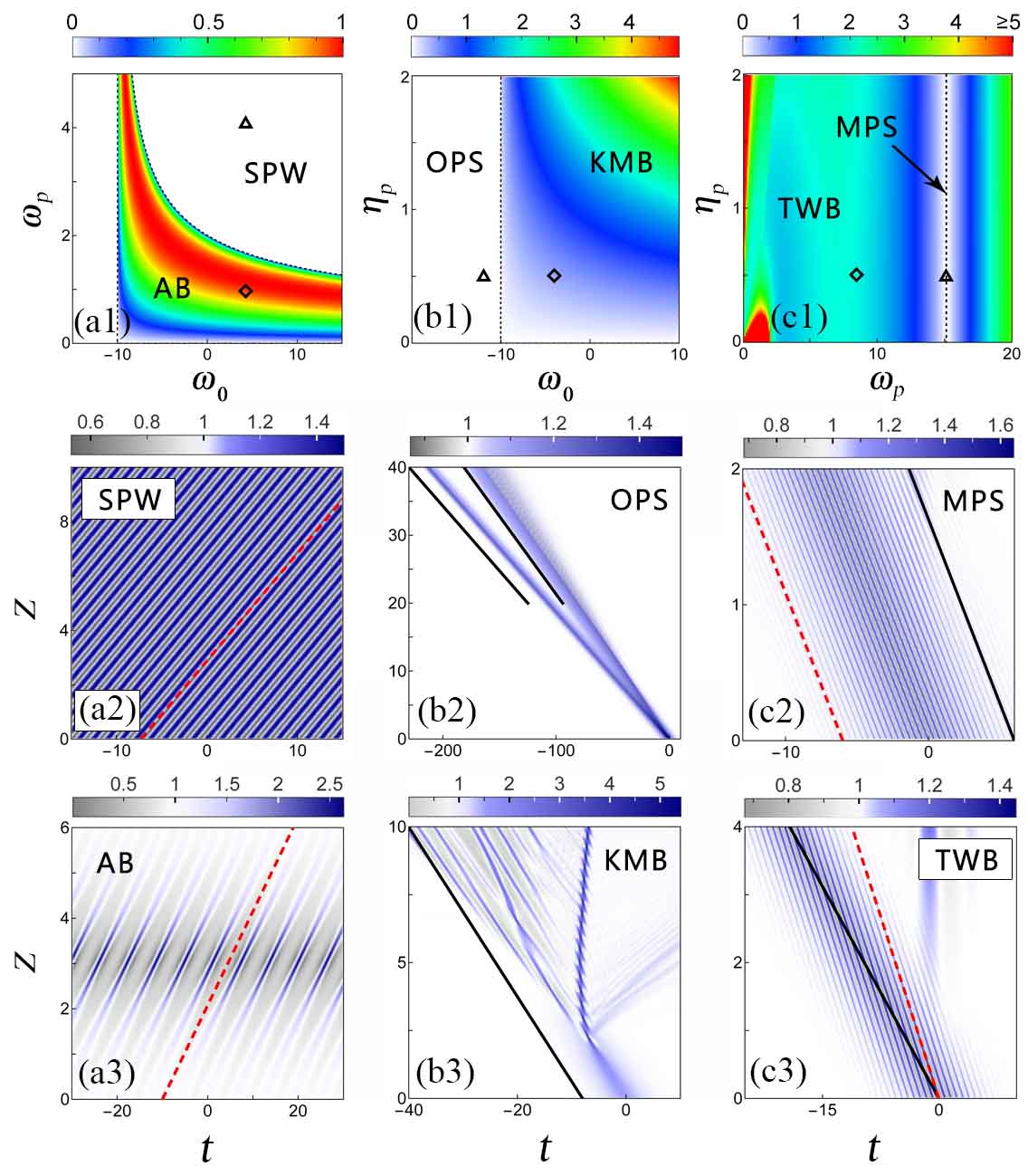}
\caption{(Color online) (a1) $|G^{(\rm II)}_{+}|$ value distribution on $\omega_0$-$\omega_p$ plane with $\eta_p=0$, where white and colored regions severally denote SPW and AB generations. (b1) $|K^{(\rm II)}_{+}|$ value distribution on $\omega_0$-$\eta_p$ plane with $\omega_p=0$, where white and colored regions severally denote OPS and KMB generations. (c1) $|V^{(\rm II)}_{+}-\Lambda^{(\rm II)}_{+}|$ value distribution on $\omega_p$-$\eta_p$ plane with $\omega_0=0$, where white and colored regions denote MPS and TWB generations. The middle and lower plots illustrate amplitude evolution for generating (a2) SPW, (a3) AB, (b2) OPS, (b3) KMB, (c2) MPS, and (c3) TWB. Their initial parameters are set as the coordinates of triangles or diamonds in different regions of (a1), (b1), and (c1). Black solid and red dashed lines represent the predicted propagating velocities of localized envelope and fringe, respectively. All the waves are generated successfully, except the KMB. The splitting in OPS generation is related to the non-degeneracy of modes and time signs. Other parameters are $\beta_3=0.1$ and $a_0=1$. }
\label{pic-nlse3}
\end{figure*}

When $\beta_3=0$, Eq. (\ref{eq-model}) is the NLS system without the third-order dispersion.
Since this system is integrable, many exact solutions describing fundamental nonlinear waves were presented \cite{Zakharov-1972,Kuznetsov-Ma,Peregrine-1983,Akhmediev-1986,Tajiri-1998}.
Among the six kinds of fundamental waves mentioned above, only AB, KMB, and TWB solutions exist in NLS systems.
It indicates that one can use a temporal profile as initial condition to generate perfect them.
Unlike it, what we focus on in this paper is how to controllably generate them from an initial condition with a general form, which is more convenient to prepare.
In our analysis on the NLS system, the expressions of $K_{\pm}$ and $G_{\pm}$ are found to be wonderfully equivalent to the coefficients describing periodicity and localization in $z$ direction given by the exact solutions of waves (which is also found in some other integrable models).
This equivalence may be related to the functional form of perturbing wave solution we assume, as it can be considered as an approximative solution of initial stage in the NLS model.

Considering that many waves in the NLS model have been excited in real fibers, we focus on the system with third-order dispersion,  which admits more kinds of wave generations.
Without loss of generality, the coefficient $\beta_3$ is set as $0.1$, and the background amplitude is $a_0=1$.
We take the mode II when $t\rightarrow +\infty$ as an example to discuss the wave generations from three types of initial perturbations.

Firstly, a purely periodic initial perturbation (with $\omega_p\neq 0$ and $\eta_p=0$) is used to generate SPW or AB.
When $\eta_p=0$, based on Eq. (\ref{eq-etg}), one can obtain
\begin{equation}
\begin{split}
\label{eq-gii}
G^{(\rm II)}_{+}=\frac{1}{2}{\rm Re}\Big[\sqrt{\omega_p^2(1+\beta_3\omega_0)
[4a_0^2-\omega_p^2(1+\beta_3\omega_0)]}\Big].
\end{split}
\end{equation}
The value of $|G^{(\rm II)}_{+}|$ is distributed on the $\omega_0$-$\omega$ plane to distinguish the generations of AB and SPW, as shown in Fig. \ref{pic-nlse3} (a1).
The white [$|G^{(\rm II)}_{+}|=0$] and colored [$|G^{(\rm II)}_{+}|\neq 0$] regions denote the generations of SPW and AB, respectively.
We set the parameters as the coordinates of triangle, namely, $(\omega_0, \omega_p)=(4, 4)$.
Its amplitude evolution is shown in Fig. \ref{pic-nlse3} (a2).
A SPW is generated perfectly, and its velocity of fringe matches well with the one predicted by the limit value of Eq. (\ref{eq-v}) (denoted by the red dashed line).
Then, we set $(\omega_0, \omega_p)=(4,1)$, which are the coordinates of diamond in Fig. \ref{pic-nlse3} (a1).
Its amplitude evolution is shown in Fig. \ref{pic-nlse3} (a3), and an AB is generated as expected.
Its numerical velocity of fringe also has a good agreement with the one we predict.

Secondly, we consider a purely localized initial perturbation (with $\omega_p=0$ and $\eta_p\neq 0$) to generate OPS or KMB.
When $\omega_p=0$, based on Eq. (\ref{eq-omk}), one can obtain
\begin{equation}
\begin{split}
\label{eq-kii}
K^{(\rm II)}_{+}=\frac{1}{2}{\rm Im}\Big[\sqrt{-\eta_p^2(1+\beta_3\omega_0)
[4a_0^2+\eta_p^2(1+\beta_3\omega_0)]}\Big].
\end{split}
\end{equation}
The distribution of $|K^{(\rm II)}_{+}|$ on $\omega_0$-$\eta_p$ plane is shown in Fig. \ref{pic-nlse3} (b1) for their generation conditions.
Unlike the NLS system, the consideration of third-order dispersion admits the OPS generation, which is denoted by the grey region [with $|K^{(\rm II)}_{+}|=0$].
The colored region [with $|K^{(\rm II)}_{+}|\neq 0$] corresponds to KMB generation.
For a purely localized initial perturbation, its initial condition (\ref{eq-ini0p}) becomes
\begin{equation}
\begin{split}
\label{eq-iniloc}
\psi_p(0,t)=[1+(A+B){\rm sech}(\eta_p t)]\;a_0e^{i\omega_0t}.
\end{split}
\end{equation}
It indicates that the generation mode cannot be selected by the combinations of different $A$ and $B$.
We set the total amplitude $A+B$ as a modest value $0.5$ to try generating KMB or OPS.
We set $(\omega_0, \eta_p)$ as the triangle's coordinates $(-12, 0.5)$, and the amplitude evolution plot is shown in Fig. \ref{pic-nlse3} (b2).
The initial localized envelope splits into to waves with different velocities.
The left-hand wave has a stable shape in the propagation process, so we consider it as a OPS;
the right-hand wave gets wider and wider with the increasing distance, and it has a typical shape of dispersive shock wave.
Compared with our analysis result, the wave splitting indicates the generation of non-degenerate waves for different modes and time signs.
The velocities of two waves need to be predicted by $\Lambda_{+}^{(\rm II)}$ and $\Lambda_{-}^{(\rm II)}$, which are shown in Fig. \ref{pic-nlse3} (b2).
They have good agreements with the numerical evolution of the two waves.
Then, we change $(\omega_0, \eta_p)$ into the diamond's coordinates $(-4,0.5)$ in Fig. \ref{pic-nlse3} (b1).
Its amplitude evolution plot is shown in Fig. \ref{pic-nlse3} (b3).
Before the appearance of second breathing period of KMB, the spontaneous oscillation emerges and breaks the generation of KMB, which indicates that the generation of KMB is not successful.
The spontaneous oscillation always has a large influence on the generation of nonlinear waves, especially for the KMB generation.
Though some efforts have been paid on generating KMBs, generating a KMB with more than one breathing period from a nonideal initial perturbation is still an open problem.

Thirdly, a periodic-localized perturbation can be used to generate MPS or TWB when the velocity matching is satisfied or not.
Based on Eq. (\ref{eq-v}), one can derive
\begin{equation}
\begin{split}
\label{eq-vii}
V^{(\rm II)}_{+}=\omega_0+\frac{\beta_3}{6}(\omega_p^2-3\eta_p^2+3\omega_0^2)
+\frac{{\rm Im}\Big[\sqrt{-P(4a_0^2+P)}\Big]}{2\omega_p},\\
\Lambda^{(\rm II)}_{+}=\omega_0+\frac{\beta_3}{6}(3\omega_p^2-\eta_p^2+3\omega_0^2)
-\frac{{\rm Re}\Big[\sqrt{-P(4a_0^2+P)}\Big]}{2\eta_p},
\end{split}
\end{equation}
where $P=(\eta_p-i\omega_p)^2(1+\beta_3\omega_0)$.
When $\omega_0=0$, the distribution of $|V^{(\rm II)}_{+}-\Lambda^{(\rm II)}_{+}|$ on $\omega_p$-$\eta_p$ plane is shown in Fig. \ref{pic-nlse3} (c1).
The colored [$|V^{(\rm II)}_{+}-\Lambda^{(\rm II)}_{+}|>0$] and white [$|V^{(\rm II)}_{+}-\Lambda^{(\rm II)}_{+}|=0$] regions are respectively corresponding to the generation of TWB and MPS.
We set the initial parameters $(\omega_p, \eta_p)$ as the triangle's coordinates in the plot.
Its amplitude evolution plot is shown in Fig. \ref{pic-nlse3} (c2).
Despite the appearance of auto-modulation process, a MPS is generated with a negative velocity.
On the localized envelope, many fringes can be observed and it can keep stable for a long distance.
Then, we change $(\omega_p, \eta_p)$ into the coordinates of diamond in Fig. \ref{pic-nlse3} (c1).
The amplitude evolution plot is shown in Fig. \ref{pic-nlse3} (c3).
Before the auto-modulation appears, a TWB is generated and propagates with a constant velocity.
Compared with the MPS generation, the localized envelope and fringe's velocities of generated TWB are obviously different, which matches well with our prediction.
Note that Eqs. (\ref{eq-gii},\ref{eq-kii},\ref{eq-vii}) work only in the cases we consider here, so they needs a recalculation when the parameter conditions are changed.

\begin{figure}[htbp]
\centering
\includegraphics[width=84mm]{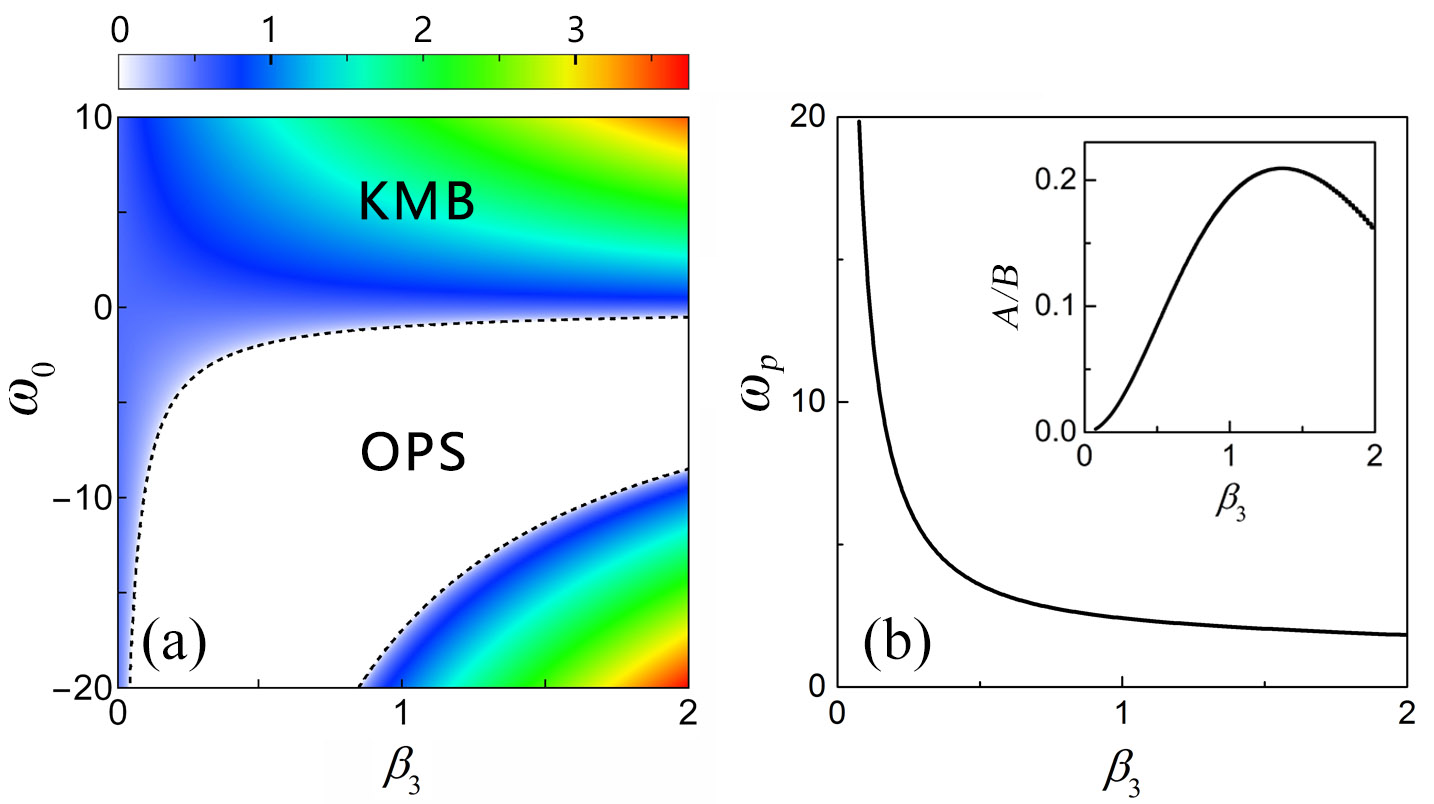}
\caption{(a) $|K_{(+ t)}^{(\rm II)}|$ value distribution on $\beta_3$-$\omega_0$ plane when $\omega_p=0$ and $\eta_p=0.5$, where the white and colored regions severally denote OPS and KMB generations. (b) Generation condition of MPS on $\beta_3$-$\omega_p$ plane when $\omega_0=0$. The inset shows the dependence of amplitude's asymmetric degree $A/B$ on $\beta_3$ for MPS generation.}
\label{pic-beta3}
\end{figure}

From the above results, the third-order dispersion has an influence on enriching the types of nonlinear waves.
The generation condition of OPS and KMB when $\omega_p=0$ and $\eta_p=0.5$ is shown in Fig. \ref{pic-beta3} (a).
With $\beta_3$ increasing, the required $\omega_0$ for OPS generation is getting higher from a negative infinite value.
When $\beta_3$ approaches $0$, the OPS generation is forbidden.
The generation condition of MPS is also shown in Fig. \ref{pic-beta3} (b) when $\omega_0=0$.
With $\beta_3$ increasing, the required $\omega_p$ for MPS generation is getting lower.
In the limit of $\beta_3\rightarrow 0$, the required $\omega_p$ has an unreachable infinite value.
This result agrees well with the case of NLS model.
Meanwhile, the asymmetric degree of initial amplitude is defined by $A/B$, and its corresponding value for MPS generation is shown in the inset.
The asymmetric degree increases from $0$ and then decreases with $\beta_3$ increasing, and it has a maximal value around $0.2$.
Its low asymmetric degree indicates the amplitude of component with frequency $\omega_p$ is far less than the one with frequency $-\omega_p$.
When $\beta_3$ is close to $0$, one can obtain $A\rightarrow 0$, which implies that the required perturbation for MPS generation has quasi-single frequency.

\section{Experimental feasibility of multi- and one-peak soliton excitations}

\begin{figure}[htbp]
\centering
\includegraphics[width=86mm]{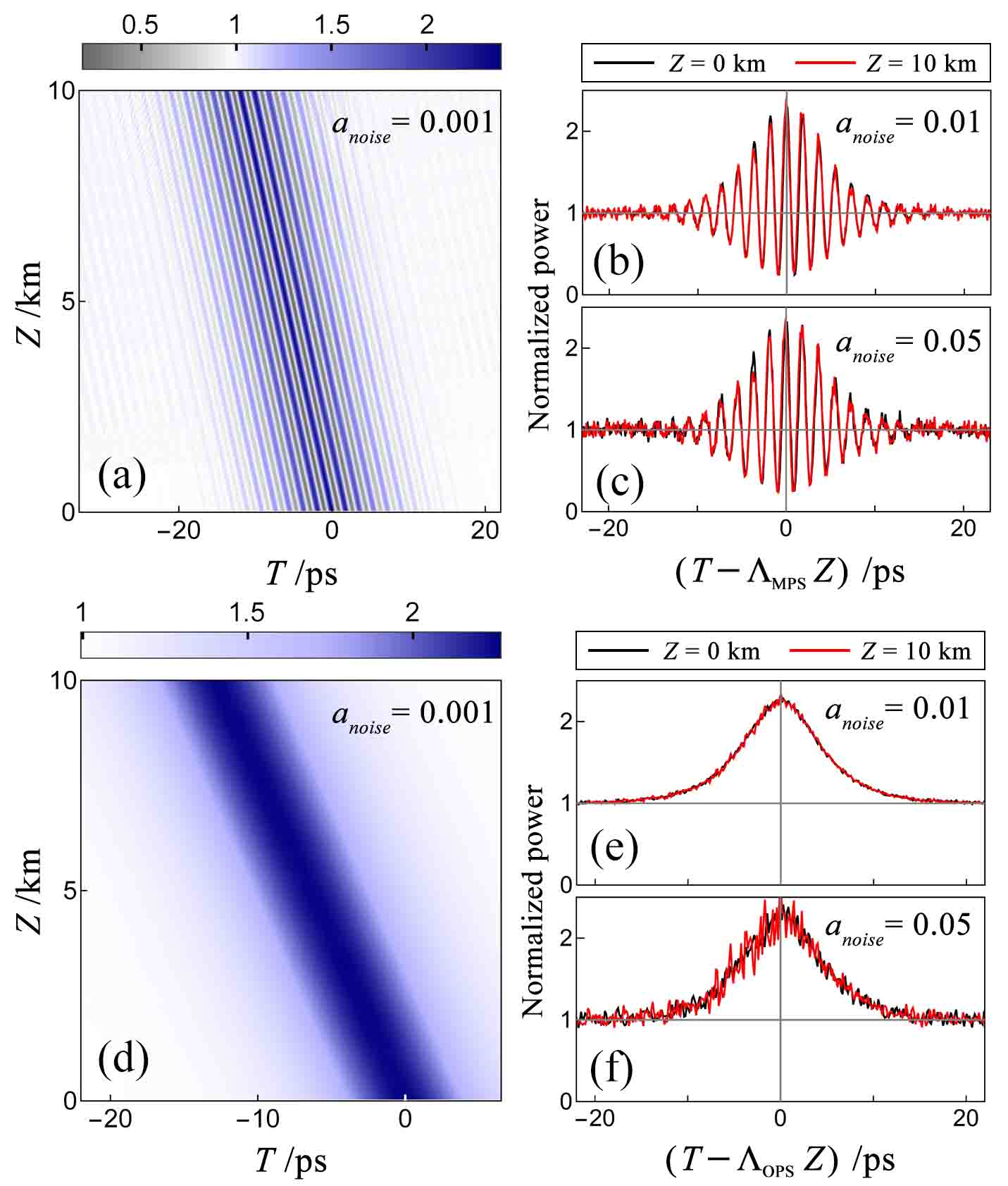}
\caption{(Color online) (a) Power evolution plot of MPS generation when $a_{noise}=0.001$. (b) Its power profile at $Z=0\,{\rm km}$ (black curve) and $Z=10\,{\rm km}$ (red curve) when $a_{noise}=0.01$. (c) The same to (b) except $a_{noise}=0.05$. (d) Power evolution plot of OPS generation when $a_{noise}=0.001$. (e) Its power profile at $Z=0\,{\rm km}$ (black curve) and $Z=10\,{\rm km}$ (red curve) when $a_{noise}=0.01$. (f) The same to (e) except $a_{noise}=0.05$. The power mentioned here is the normalized power, $|A(Z,T)|^2/[P_0\exp(-\alpha Z)]$. $\Lambda_{\rm MPS}$ and $\Lambda_{\rm OPS}$ are the envelope velocities of MPS and OPS, respectively.}
\label{pic-exp}
\end{figure}

Now, we focus on the experimental feasibility for exciting multi- and one- peak solitons.
By considering $A=\sqrt{P_0} \psi$, $T=\sqrt{|\beta^{(2)}|/(\gamma P_0)}t$, and $Z=(\gamma P_0)^{-1}z$, the model (\ref{eq-model}) can be transformed into a dimensional model.
When the fiber loss is considered, the model is
\begin{eqnarray}
\label{eq-modela}
iA_Z-\frac{\beta^{(2)}}{2}A_{TT} -\frac{i\beta^{(3)}}{6}A_{TTT}+\gamma |A|^2A+i\frac{\alpha}{2} A=0,
\end{eqnarray}
which describes the optical wave evolution in real fibers.
Here, $P_0$, $\gamma$, $\beta^{(2)}$, $\beta^{(3)}$, and $\alpha$ scale respectively the input power, the nonlinearity, the group velocity dispersion, the third-order dispersion, and the fiber loss.
The relation between the two coefficients of third-order dispersion is $\beta_3=\sqrt{\gamma P_0/|\beta^{(2)}|^3} \beta^{(3)}$.
We set $\beta^{(2)}=-0.5065\,{\rm ps^2/km}$, $\beta^{(3)}=0.1\,{\rm ps^3/km}$, $\gamma=1.3\,{\rm W^{-1}km^{-1}}$, $\alpha=4.2\times10^{-3}\,{\rm km^{-1}}$, and the input power $P_0=0.1\,{\rm W}$.
The initial condition is still the continuous wave perturbed by a sech-type envelope with two-frequency carrier waves,
\begin{equation}
\begin{split}
\label{eq-ini-exp}
A_p(0,T)=\sqrt{P_0}e^{i\Omega_0T}[1+(A_1e^{i\Omega_p T}+A_2e^{-i\Omega_p T}){\rm{sech}} (H_p T)].
\end{split}
\end{equation}
The phase of continuous wave is modulated periodically by the frequency $\Omega_0$.
The relative periodicity and localization of perturbation are scaled by $\Omega_p$ and $H_p$, respectively.
When we consider the influence of random noise on wave excitations, the initial condition becomes
\begin{equation}
\begin{split}
\label{eq-noise}
A_{noise}(0,T)=A_p(0,T)(1+a_{noise}{\rm Random}[-1,1]),
\end{split}
\end{equation}
where $a_{noise}$ is the relative amplitude of random noise, and the function, ${\rm Random}[a,b]$, can produce a random value between $a$ and $b$ at every numerical time point.

For the excitation of MPS, we set $\Omega_0=-461.22\,{\rm GHz}$, $\Omega_p=520.97\,{\rm GHz}$, $H_p=38.44\,{\rm GHz}$, $A_1=0.03$, and $A_2=0.5$, based on the above analysis.
When $a_{noise}=0.001$, the power evolution plot is shown in Fig. \ref{pic-exp} (a).
Its color scale denotes the normalized power, $|A(Z,T)|^2/[P_0\exp(-\alpha Z)]$.
One can see that a MPS propagating stably is excited.
When the noise amplitude is increased, its profiles at the initial and final distances are compared in Fig. \ref{pic-exp} (b) and (c).
The good agreements between them indicates that the MPS keeps robust against increased noises.
Then, we try the excitation of OPS by setting $\Omega_0=-768.71\,{\rm GHz}$, $\Omega_p=0\,{\rm GHz}$, $H_p=38.44\,{\rm GHz}$, and $A_1+A_2=0.5$.
This set is locating on the critical line between OPS and KMB regions in Fig. \ref{pic-nlse3} (b1), which avoids the splitting of initial perturbation.
Its power evolution plot is shown in Fig. \ref{pic-exp} (d) when $a_{noise}=0.001$.
A OPS propagates stably with a constant velocity.
Its initial and final profiles are compared under different $a_{noise}$ in Fig. \ref{pic-exp} (e) and (f).
When $a_{noise}=0.01$, small fluctuations appear, and the final profile agrees well with the initial one.
When $a_{noise}=0.05$, though the fluctuation in the final profile is amplified, the profile of soliton is still kept well.
These results indicates that a MPS has a better robustness than a OPS when a noise is considered.

\section{Conclusion}

In conclusion, we numerically generate six kinds of nonlinear waves, and control their type, velocities, periodicity, and localization, in the NLS model with third-order dispersion.
The generation condition of these waves is illustrated by means of the modified linear stability analysis method.
They are generated from a cosine-type initial perturbation with a sech-type envelope on continuous waves.
The velocities of their localized envelope and fringes are predicted successfully.
Besides, the introduction of third-order dispersion is found to admit the generation of OPS and MPS so riches the kinds of waves.
The results can provide the guidance for wave generations in real optical fibers and other nonlinear systems, like Bose-Einstein condensates, water, and plasmas.

\section*{Acknowledgement}
%The authors thank Dr. L. Duan for his helpful discussions.
This work was supported by National Natural Science Foundation of China (Contact No. 11875220; 11947301).

\end{document}